\documentclass[conference]{IEEEtran}
\IEEEoverridecommandlockouts
\usepackage{cite}
\usepackage{float}
\usepackage{amsmath,amssymb,amsfonts}
\usepackage{algorithmic}
\usepackage{graphicx}
\usepackage{textcomp}
\usepackage{xcolor}
\usepackage{svg}
\usepackage{cuted} 
\usepackage{multirow}  
\usepackage{booktabs}  
\usepackage{capt-of} 
\usepackage{placeins}
\usepackage{url}
\usepackage{tabularx}
\usepackage{array}

\def\BibTeX{{\rm B\kern-.05em{\sc i\kern-.025em b}\kern-.08em
    T\kern-.1667em\lower.7ex\hbox{E}\kern-.125emX}}
\begin{document}
\bstctlcite{BSTcontrol}
\title{Performance Comparison of IBN orchestration using LLM and SLMs\\

}

\author{
\IEEEauthorblockN{
Wai Lwin Phone\IEEEauthorrefmark{1},
Brahim El Boudani\IEEEauthorrefmark{1},
Tasos Dagiuklas\IEEEauthorrefmark{1},
Saptarshi Ghosh\IEEEauthorrefmark{1}\IEEEauthorrefmark{2}
}
\IEEEauthorblockA{\IEEEauthorrefmark{1}Dept. of Computer Science \& Digital Technologies, London South Bank University, London, UK}
\IEEEauthorblockA{\IEEEauthorrefmark{2}Digital Catapult, London, UK}
\IEEEauthorblockA{\IEEEauthorrefmark{1}s4220393@lsbu.ac.uk, elboudani@lsbu.ac.uk, tdagiuklas@lsbu.ac.uk, ghoshs5@lsbu.ac.uk}
\IEEEauthorblockA{\IEEEauthorrefmark{2}saptarshi.ghosh@digicatapult.org.uk}
}

\maketitle

\begin{abstract} The evolution of both 5G and 6G networks is driving the advancement of fully autonomous network management, placing Intent-Based Networking at the centre of this transformation. This paper introduces a novel framework for 5G and 6G IBN orchestration that leverages a stateful, hierarchical multi-agent architecture to achieve full automation using both SLMs and LLMs. Both models have been evaluated for translation accuracy using metrics such as BLEU, METEOR, and ROUGE-L, as well as computational complexity. Experimental results show that both models exhibit similar accuracy. However, result shows that SLMs can improve the overall completion speed of the IBN lifecycle by 20\%.\end{abstract}

\begin{IEEEkeywords}
IBN, Orchestration, SLM, LLM
\end{IEEEkeywords}

\section{Introduction}
The rapid evolution of mobile network technologies, particularly driven by 5G, is enabling diverse application verticals. These include enhanced Mobile Broadband (eMBB), Ultra-Reliable Low-Latency Communications (URLLC), and massive Machine-Type Communications (mMTC). Meanwhile, there is increasing demand for concurrent services delivered by over-the-top (OTT) platforms \cite{Brodimas2025}. The expanding, diverse customer base requires new levels of network management. Capabilities such as service customisation and resource allocation are essential to handle complex, heterogeneous service requirements while optimising continuous capacity utilisation \cite{Brodimas2025,WuXXXX}. However, traditional network management in (industrial) IoT and critical infrastructure relies mostly on rule-based, infrequent configuration changes \cite{Viticchie2025}. This rigidity in transport networks is common in larger telecommunication systems. It is often justified by the need for strict security and regulatory compliance, such as meeting FISMA and NIST standards. This rigidity severely limits the ability to adapt to dynamicity \cite{Viticchie2025,Ingole2024}. To overcome these limits, the network engineering field has embraced \textit{softwarization}. This trend gave rise to Software-Defined Networking (SDN) and its derivative, Intent-Based Networking (IBN) \cite{Brodimas2025}. IBN represents a shift by simplifying network management, translating high-level user goals or intents into actionable network configurations. This reduces the need for detailed manual steps. Highly programmable networks now use RESTful API-based interactions to streamline configuration and deployment \cite{Brodimas2025,Ingole2024,Sacco2025}. Building on programmability and abstraction, recent network management research turns to advanced Artificial Intelligence (AI) solutions. These help automate and enhance orchestration \cite{WuXXXX}. Generative AI (GenAI) and Large Language Models (LLMs) now present a chance to transform natural language requirements into textual network configurations. This enables sophisticated IBN deployments \cite{Brodimas2025,WuXXXX}. LLMs excel at reasoning, context, and generating human-like text. This makes them powerful tools for interpreting user intent \cite{Wang2024,Acharya2025}. Their capability has driven interest in LLM-based network orchestration in research \cite{Brodimas2025,WuXXXX}. However, the stochastic nature of LLMs makes them difficult to train. They can struggle to deliver consistent, robust outcomes needed for reliable control and infrastructure provisioning \cite{Raza2025}. To address this reliability gap, the field is now moving toward Agentic AI \cite{Brodimas2025}. Agentic AI uses distributed intelligence, deploying constrained, autonomous learners called agents. In multi-agent systems, specialised agents decompose intents, plan actions, and execute tasks reliably using external tools. This paper introduces a new hierarchical multi-agent framework for autonomous 5G/6G IBN orchestration. It integrates specialised LoRA-tuned Small Language Models (SLMs) and LLMs. The paper also presents an experimental evaluation of their accuracy and time complexity. Results show that SLMs achieve accuracy comparable to LLMs while improving IBN lifecycle completion speed by about 20\%, providing an efficient route for network automation.

The remainder of this paper is structured as follows. Section II overviews Agentic-AI in network automation, focusing on SLM and LLM contributions. Section III presents our stateful, multi-agent architecture and the roles of the agents. Section IV details the experimental setup, SLM fine-tuning, and agent performance evaluation. Section V concludes and outlines future work.

\section{State-of-the-Art in Agentic-AI for Network Automation}
IBN enhanced by Agentic AI introduces distributed multi-agent systems where LLM-powered agents decompose user intents and adapt configurations via Infrastructure-as-Code (IaC) \cite{Brodimas2025}. Collaborative agents handle negotiation and conflict resolution across business, service, and network planes in multi-tenant 6G networks \cite{Chatzistefanidis2024}, while domain-adapted LLMs like ChatNet bridge natural and network-specific languages, integrating external tools for capacity planning and policy translation \cite{Huang2025}. These systems provide network-embodied intelligence for automated Q\&A, code generation, incident management, and AI-driven monitoring, planning, deployment, and optimisation across mobile, vehicular, cloud, and edge networks \cite{Boateng2025}.

LLMs are proficient in text generation and complex reasoning but suffer from large parameter sizes and computational demands. SLMs address these limitations by offering lower latency and reduced costs, making them ideal for environments such as edge devices and mobile platforms \cite{wang2024comprehensive}. While Knowledge Distillation (KD) is a common approach for creating SLMs by training a smaller model to replicate the "soft label" outputs of a larger "teacher" LLM \cite{gu2023minillm}, it requires significant GPU computing power and risks propagating the teacher's inherent biases and hallucinations. In contrast, we train the SLMs on an actual dataset rather than an approximation. This direct specialisation enables SLMs to achieve higher accuracy on their target task \cite{hu2022lora}.

In the networking domain, LLMs are transforming the automation of 5G and 6G network management. Traditional AI/ML networking solutions for traffic prediction and anomaly detection lack generalisation and struggle with unstructured data such as logs and telemetry \cite{Huang2025,Long2025}. Recent surveys present a unified LLM workflow encompassing task definition, data representation, prompt engineering, model evolution, tool integration, and validation \cite{Liu2025}. Benchmarks like NetConfEval demonstrate LLMs' capability to translate high-level requirements into network specifications, generate routing algorithms, and configure devices with high accuracy \cite{Wang2024}.

Although prior work demonstrates the potential of LLM-powered agents for network automation, it mainly emphasises intent translation, prompt-driven configuration, or centralised orchestration. This doesn't define a standardised framework for structured inter-agent collaboration or context-aware knowledge exchange across distributed domains. Moreover, the reliance on large general-purpose LLMs introduces latency and scalability limitations in real-time networking environments.

To address these gaps, a distributed multi-agent architecture is proposed, built on a dedicated Agent-to-Agent (A2A) protocol and a Model Context Protocol (MCP), enabling modular, asynchronous, and context-aware coordination among specialised agents. Unlike existing approaches that rely on knowledge distillation or prompted LLMs, the framework leverages SLMs. These SLM-based agents collaborate within a \textit{Routing-as-a-Service} (RaaS) \cite{iraas} to provide a unified control plane across heterogeneous SDN and legacy infrastructures. The combination of the A2A protocol, MCP-based context management, specialised SLM agents, and RaaS integration advances the state of the art by enabling intent-based network automation.

\section{Proposed Agentic-AI Network Architecture}
\begin{figure}[t]
  \centering
  \includegraphics[width=\linewidth]{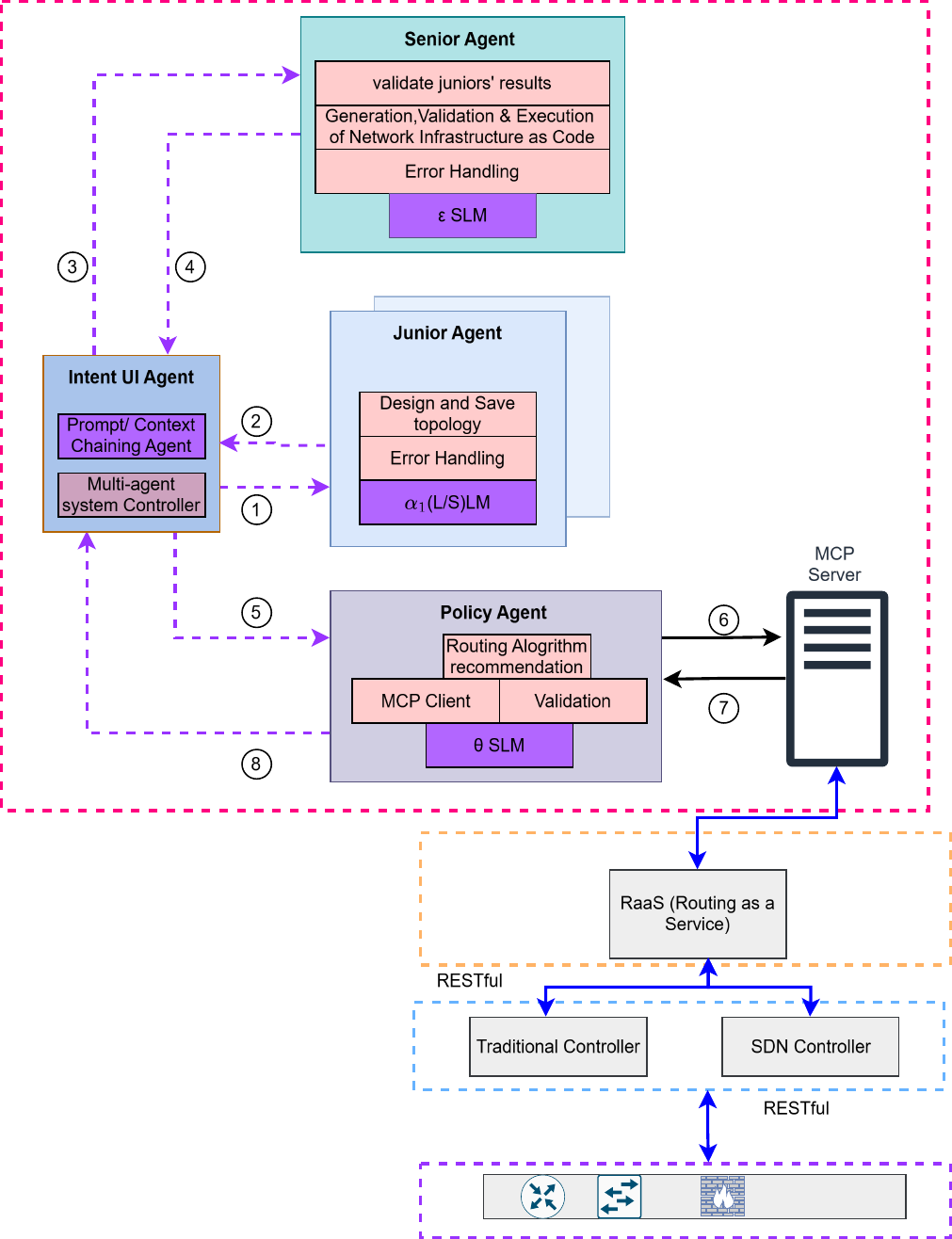} 
  \caption{Proposed IBN Orchestration Architecture.}
  \label{fig:diagram}
\end{figure}

\subsection{ Multi-Agent Architecture}
Stateful hierarchical multi-agent architectures have emerged as a common and effective design pattern, favouring a coordinated team of specialised agents over a single model. This structure enables a clear division of responsibilities and allows each agent to apply domain-specific expertise \cite{Viticchie2025, Brodimas2025}.

In Figure \ref{fig:diagram} the Intent UI Agent serves as an entry point for interacting with both users and available agents. It contains a Prompt Chaining Multi-Agent System Controller that retrieves user intentions via an interface and initiates the initial prompt-chaining sequence. It sends the user's request in parallel to Junior Agents to build out topologies and services, and it loops back indirectly for checks and validation with the Senior Agent. This validated output is forwarded to the Policy Agent, which then suggests a routing protocol based on the user's intent. The MCP client for policy-driven validation can use either SLMs or LLMs and interface with the MCP Server to deploy configurations via RaaS element, data plane controllers, and associated protocols.

The Junior Agents are responsible for interpreting user requirements to generate corresponding network topology and service configurations. Utilising either SLMs or LLMs, these agents handle requests, synthesise network configurations, and maintain persistent storage of relevant information. Upon completion, their output is relayed through the Intent UI Agent to a Senior Agent for validation.

The use of two Junior Agents implements a comparison-based fault detection model inspired by Dual-Modular Redundancy (DMR) \cite{koren2020fault}. Rather than masking faults through majority voting, this design focuses on error detection, well-suited to deterministic network engineering tasks where correct configurations are clearly defined. Matching outputs from both agents indicates high confidence, while any divergence signals uncertainty and triggers escalation.

Conflicts are resolved by a Senior Agent, who acts as a reasoning comparator. It evaluates competing configurations against Policy Agent constraints and selects the optimal solution without requiring a third voting entity. The Senior Agent then translates the validated design into deterministic infrastructure code.

The emulation environment serves as the final validator. If instantiation fails, it provides a definitive failure signal, effectively acting as the ultimate redundant partner. Through validation, error handling, and code generation, the Senior Agent ensures the correctness and reliability of the final network configuration.

Policy Agent is assigned to analyse network topology for configuration validation and to generate intelligent routing recommendations based on real-time network state data obtained from the MCP Server. The MCP Server maintains connectivity with both RaaS components and underlying traditional and SDN controllers.
Finally, the system addresses the common reliance on Human-in-the-Loop (HITL) strategies often used for security and refinement in AI systems \cite{rizos2025concept, Brodimas2025}. While acknowledging the value of HITL for creative or sensitive operations \cite{viticchie2025ai}, our approach mitigates the bottlenecks created by human intervention in repetitive tasks.

The Senior Agent is responsible for ensuring the correctness and reliability of network infrastructure configurations. It validates junior agents' topology-generation results and generates appropriate topology weights to guide network optimisation decisions. The Senior Agent implements error handling and generates, validates, and deploys network infrastructure as executable code.

RaaS provides unified routing across SDN and traditional IP domains via a distributed control plane. Domain-specific controllers, SDN via OpenFlow, and IP routers via SSH, aggregate local topologies into a global network graph. The RaaS server applies user-defined routing logic such as SPF (RFC 2328), DUAL (RFC 7868) and metrics like load, reliability, and latency. Inter-domain topologies are merged through a central pseudo-node, and the resulting paths can be converted into SDN flow entries or IP Route-Maps for traditional routers.

\subsection{Message Sequence}
When a user submits an intent request through the interface, the Intent UI Agent triggers the process by passing the request to Junior Agent 1 and Junior Agent 2. Working together, the system generates an initial set of topology recommendations. Next, the system proceeds to the validation phase. The Senior Agent initiates validation with a configurable retry mechanism (up to three attempts). This verification process is performed for a maximum of three iterative attempts, a configuration supported by research showing this number balances accuracy improvements with resource utilisation \cite{Wei2022}. If validation passes, the workflow continues to the Network Policy System component. If it fails, the Junior Agents generate alternative topology recommendations, and the same validation process is repeated until a valid configuration is found or retries are exhausted.

Once the topology passes initial validation, the system proceeds to network instantiation. Here, configuration scripts are generated and executed to deploy the desired topology. A second validation cycle then takes place, also with up to three attempts. This phase involves collecting live topology data, computing graph weights, integrating them with the topology, and analysing routing algorithm performance. The outcome determines whether the deployment is valid or requires further modifications. Throughout these stages, the system maintains detailed logs that store all outputs and acknowledgements. This helps trace every step in the topology-generation, validation, and policy-enforcement pipeline.

\section{Performance Evaluation}
\subsection{Experimentation Setup}
\begin{figure}[t]
  \centering
  \includegraphics[width=\linewidth, height=0.9\textheight, keepaspectratio]{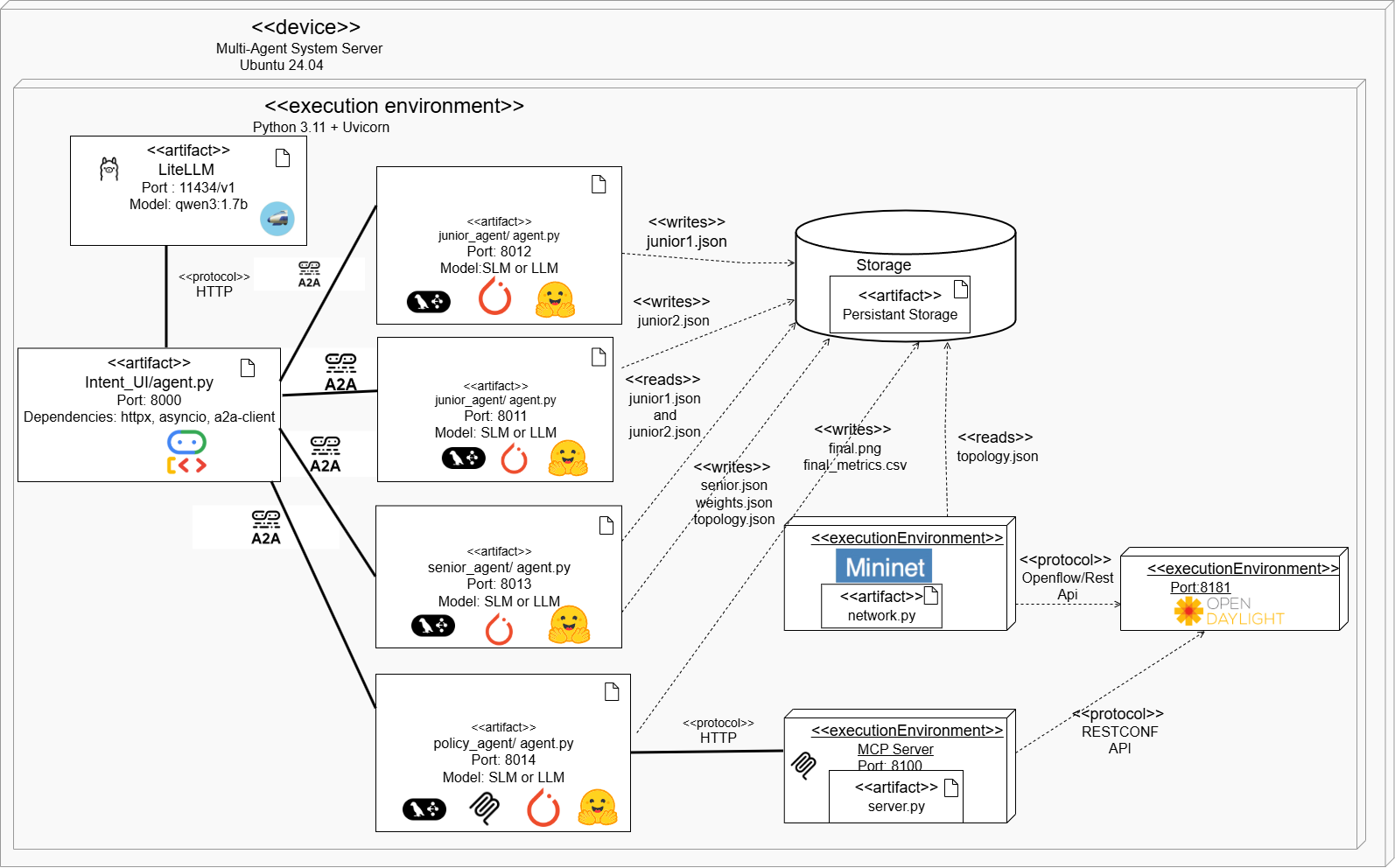}
  \caption{Deployment Diagram}
  \label{fig:diagram_wide}
\end{figure}

Figure \ref{fig:diagram_wide} illustrates the proposed multi-agent system, deployed on an Ubuntu server using Python and Uvicorn. The system is centred around an Intent UI agent that orchestrates four specialised agents: two Junior agents, one Senior agent, and one Policy agent. The central UI agent leverages the Qwen3:1.7B served via Ollama and accessed through LiteLLM on TCP port 8000. The specialised agents utilise a combination of SLMs or LLMs, depending on task complexity and performance requirements.

The system follows a distributed deployment architecture, with each agent operating on a distinct port: Junior agents on 8011 and 8012, the Senior agent on 8013, and the Policy agent on 8014. This design allows agents to be hosted on separate physical or virtual servers, enabling horizontal scaling. Agents communicate asynchronously over HTTP using the A2A standard protocol and interact with external services, including a Mininet simulation and an MCP server. All data, including agent outputs and network configurations, are stored and managed in JSON format.

The experimental setup runs on a GPU cluster comprising 4x GeForce RTX 2080 Ti 11 GB GDDR6 coupled with an Intel(R) Core(TM) i9-9820X CPU @ 3.30GHz processor and 32 GB of system memory. 


\begin{table*}[t]
\caption{Scenarios (LLMs and SLMs and network topologies)}
\label{tab:tab1}
\centering
\scriptsize
\setlength{\tabcolsep}{5pt}
\renewcommand{\arraystretch}{1.15}

\begin{tabularx}{\textwidth}{@{}
>{\raggedright\arraybackslash}p{2.8cm}
>{\raggedright\arraybackslash}p{2.4cm}
>{\raggedright\arraybackslash}p{3.0cm}
X
>{\raggedright\arraybackslash}p{2.3cm}
>{\raggedright\arraybackslash}p{2.0cm}
@{}}
\hline\hline
\textbf{Model} & \textbf{Agent Type} & \textbf{Role} & \textbf{Topology Options} & \textbf{5G Service Types} & \textbf{Routing Strategy} \\
\hline

\multirow{4}{*}{\textbf{TinyLlama-1.1B (SLM)}} 
& Junior Agent 1 & Topology \& Service Analysis & Full-mesh, Partial-mesh, Hub-and-spoke & eMBB, URLLC, MMTC & -- \\
& Junior Agent 2 & Topology \& Service Analysis & Full-mesh, Partial-mesh, Hub-and-spoke & eMBB, URLLC, MMTC & -- \\
& Senior Agent    & Generate topology \& weights & -- & -- & -- \\
& Policy Agent    & Decide routing strategy      & -- & -- & OSPF or Dual \\
\hline

\multirow{4}{*}{\textbf{SmolLM-1.7B (SLM)}} 
& Junior Agent 1 & Topology \& Service Analysis & Full-mesh, Partial-mesh, Hub-and-spoke & eMBB, URLLC, MMTC & -- \\
& Junior Agent 2 & Topology \& Service Analysis & Full-mesh, Partial-mesh, Hub-and-spoke & eMBB, URLLC, MMTC & -- \\
& Senior Agent    & Generate topology \& weights & -- & -- & -- \\
& Policy Agent    & Decide routing strategy      & -- & -- & OSPF or Dual \\
\hline

\multirow{4}{*}{\textbf{gpt-5-nano}} 
& Junior Agent 1 & Topology \& Service Analysis & Full-mesh, Partial-mesh, Hub-and-spoke & eMBB, URLLC, MMTC & -- \\
& Junior Agent 2 & Topology \& Service Analysis & Full-mesh, Partial-mesh, Hub-and-spoke & eMBB, URLLC, MMTC & -- \\
& Senior Agent    & Generate topology \& weights & -- & -- & -- \\
& Policy Agent    & Decide routing strategy      & -- & -- & OSPF or Dual \\
\hline

\multirow{4}{*}{\textbf{mistral-small 3.1:24b}} 
& Junior Agent 1 & Topology \& Service Analysis & Full-mesh, Partial-mesh, Hub-and-spoke & eMBB, URLLC, MMTC & -- \\
& Junior Agent 2 & Topology \& Service Analysis & Full-mesh, Partial-mesh, Hub-and-spoke & eMBB, URLLC, MMTC & -- \\
& Senior Agent    & Generate topology \& weights & -- & -- & -- \\
& Policy Agent    & Decide routing strategy      & -- & -- & OSPF or Dual \\
\hline\hline

\end{tabularx}
\end{table*}

Table \ref{tab:tab1} illustrates the network scenarios evaluated. Quality-of-service policy routing requirements are enforced for applications such as eMBB and URLLC, while mMTC applications may use simpler connectivity-based routing schemes, such as SPF.

\subsection{Fine-Tuning of Models}
To reduce the computational cost of fine-tuning SLMs, this work applies Low-Rank Adaptation (LoRA), freezing pre-trained weights and injecting trainable low-rank matrices into the Transformer to lower memory and parameter requirements \cite{hu2022lora}. TinyLlama-1.1B was selected for fine-tuning due to its strong performance among models of similar size, pretrained on 1–3 trillion tokens with optimisations like FlashAttention \cite{zhang2024tinyllama}. GPT-5-Nano and Mistral-Small(24B) were used via prompt engineering as comparative baselines. Results show that LoRA fine-tuning on TinyLlama-1.1B achieves competitive accuracy and efficiency relative to prompt-optimized LLMs.

\subsection{Performance Evaluaion Methologies}
The following subsection describes the various performance evaluation metrics we leveraged in benchmarking our models.

\subsubsection{Bilingual Evaluation Understudy (BLEU)}
To evaluate both LLM and SLMs performance, the Bilingual
Evaluation Understudy (BLEU) score has been used (Eq. \ref{eq1}), since it is a standard metric for
automatic machine translation assessment\cite{Papineni2002}. 

\begin{equation}
\mathrm{BLEU} = B_p \times \exp \left( \sum_{n=1}^{N} w_n \log p_n \right)
\label{eq1}
\end{equation}
The BLEU score evaluates translation quality by combining a Brevity Penalty ($B_p \in \mathbb{R}^{+}$), which penalises translations shorter than their references, with the geometric mean of modified $n$-gram precisions ($p_n \in \mathbb{R}$) for $n\in \{1..N\}$ and $w_n \in \mathbb{R}^{+}$ being the weights. In our experimental setup we take $N=4$. This core precision metric measures the overlap of word sequences between the candidate and reference translations, using a clipping mechanism to avoid artificially high scores from over-generated phrases. By calculating a geometric mean, BLEU ensures a balanced assessment of both adequacy (word choice via unigrams) and fluency (word order via longer $n$-grams). 

\begin{table}[t]
\caption{BLEU-2 score comparison across agents and models ($N=500$ per group). BLEU-2 is computed at the corpus level with equal unigram and bigram weights (no smoothing).}
\label{tab:bleu_comparison_single_col}
\centering
\small
\begin{tabular}{l l c c} 
\toprule
\textbf{Agent Type} & \textbf{Model} & \textbf{$\mu$} & \textbf{$\sigma$} \\ 
\midrule
\multirow{3}{*}{Senior Agent} 
 & TinyLlama-1.1B(SLM) & 0.8884 & 0.08 \\
 & mistral-small3.2:24b & 0.8240 & 0.09 \\
 & gpt-5-nano & 0.7621 & 0.10 \\
\midrule
\multirow{3}{*}{Policy Agent} 
 & TinyLlama-1.1B(SLM) & 0.8080 & 0.11 \\
 & gpt-5-nano & 0.7886 & 0.14 \\
 & mistral-small3.2:24b & 0.7226 & 0.14 \\
\midrule
\multirow{3}{*}{Junior Agent} 
 & TinyLlama-1.1B(SLM) & 0.7886 & 0.11 \\
 & gpt-5-nano & 0.7679 & 0.13 \\
 & mistral-small3.2:24b & 0.4507 & 0.34 \\
\bottomrule
\end{tabular}
\end{table}

\subsubsection{Metric for Evaluation of Translation with Explicit ORdering (METEOR)}
METEOR evaluates a candidate translation by first creating an alignment between its unigrams and those in one or more reference translations. This alignment process is flexible, accommodating not only exact word matches but also stemmed forms and synonyms. Based on this alignment, precision ($P \in \mathbb{R}^{+}$) and recall ($R \in  \mathbb{R}^{+}$) are calculated for the unigrams \cite{Banerjee2005}. The core of the METEOR score is a weighted harmonic mean of precision and recall, termed $F_{\text{mean}}$, which places a greater emphasis on recall, computed as (Eq. \ref{eq2}). It ensures that translations which successfully capture the content of the reference are favoured.

\begin{equation}
    F_{\text{mean}} = \frac{10PR}{R+9P}
    \label{eq2}
\end{equation}

To account for the fluency and grammatical correctness of the translation, a penalty term is introduced. The penalty is calculated based on the number of contiguous chunks of matched words (Eq. \ref{eq3})
\begin{equation}
    \text{Penalty} = 0.5 \times \left( \frac{C}{M} \right)^3
    \label{eq3}
\end{equation}

where $C\in \mathbb{N}$ is the number of \textit{chunks}, defined as contiguous sequences of matched words that occur in the same order for both candidate and reference sentences and $M \in \mathbb{N}$ is the total number of matched unigrams between the candidate and reference sentences. A lower number of chunks for a given number of matched words signifies better word order and thus incurs a smaller penalty. The final METEOR score is then calculated by modulating the $F_{\text{mean}}$ with this penalty (Eq. \ref{eq4}). 
\begin{equation}
    \text{Score} = F_{\text{mean}} \times (1 - \text{Penalty})
    \label{eq4}
\end{equation}
\begin{table}[t]
\centering
\caption{Meteor score comparison across agents and models ($N=500$ per group). METEOR: SENTENCE-LEVEL AVERAGE USING NLTK’S IMPLEMENTATION.}
\label{tab:meteor_scores_table_ii}
\begin{tabular}{llcc}
\toprule
\textbf{Agent Type} & \textbf{Model} & \textbf{$\mu$} & \textbf{$\sigma$} \\
\midrule
\multirow{3}{*}{Senior Agent}
 & TinyLlama-1.1B(SLM) & 0.8839 & 0.06 \\
 & mistral-small3.2:24b & 0.8748 & 0.07 \\
 & gpt-5-nano & 0.8349 & 0.06 \\
\midrule
\multirow{3}{*}{Policy Agent}
 & TinyLlama-1.1B(SLM) & 0.7077 & 0.11 \\
 & gpt-5-nano & 0.5800 & 0.23 \\
 & mistral-small3.2:24b & 0.4111 & 0.33 \\
\midrule
\multirow{3}{*}{Junior Agent}
 & TinyLlama-1.1B(SLM) & 0.7665 & 0.09 \\
 & mistral-small3.2:24b & 0.6873 & 0.19 \\
 & gpt-5-nano & 0.6008 & 0.22 \\
\bottomrule
\end{tabular}
\end{table}

\subsubsection{Recall-Oriented Understudy for Gisting Evaluation(ROUGE)}
For evaluating the performance of the text-generation agent,   the ROUGE-L metric has been used \cite{Lin2004}, which is grounded in the concept of the Longest Common Subsequence $R_{lcs}$ (Eq. \ref{eq5}). This metric measures the longest sequence of words that appears in both the machine-generated text (Candidate, $C \in \mathbb{N}$) and the human-written ground truth (Reference, $R \in \mathbb{N}$), while preserving their relative order but not requiring them to be consecutive. The ROUGE-L recall quantifies how much of the reference's content and structure is successfully captured in the candidate text.
\begin{equation}
R_{\text{lcs}} = \frac{\text{LCS}(C,R)}{|R|}
\label{eq5}
\end{equation} 
ROUGE-L has been selected, because it offers a balance of flexibility and structural assessment compared to $n$-gram-based metrics in ROUGE. While $n$-gram metrics are insensitive to word order or overly strict, ROUGE-L effectively rewards sentence-level coherence. This characteristic is particularly vital for the agent(s), which must generate justifications that are semantically sound and logically structured, rather than just matching specific phrases.

\begin{table}[t]
\centering
\caption{ROUGE-L score comparison across agents and models ($N = 500$ per group), with standard deviation. ROUGE-L: Average F-measure computed with stemming enabled.}
\begin{tabular}{l l c c}
\toprule 
\textbf{Agent Type} & \textbf{Model} & \textbf{$\mu$} & \textbf{$\sigma$} \\
\midrule 
\multirow{3}{*}{Senior Agent} 
& TinyLlama-1.1B(SLM) & 0.8810 & 0.09\\
& mistral-small3.2:24b & 0.8578 & 0.08\\
& gpt-5-nano & 0.7769 & 0.12\\
\midrule 
\multirow{3}{*}{Policy Agent} 
& TinyLlama-1.1B(SLM) & 0.6987 & 0.17\\
& gpt-5-nano & 0.6600 & 0.18\\
& mistral-small3.2:24b & 0.5422 &  0.25\\
\midrule 
\multirow{3}{*}{Junior Agent} 
& TinyLlama-1.1B(SLM) & 0.8089 & 0.08\\
& mistral-small3.2:24b & 0.6647 & 0.16 \\
& gpt-5-nano & 0.6017 & 0.18\\
\bottomrule 
\end{tabular}
\end{table}

The results show that lightweight,models, when orchestrated properly, can deliver reliable and deterministic network automation. Unlike existing approaches that depend mainly on large, general-purpose LLMs with their high computational costs and unpredictable behavior our work demonstrates how domain-adapted SLMs can collaborate within a multi-agent system to provide a cost effective solutions for IBN.

\subsection{Time Distribution}
\begin{table}[t]
\centering
\caption{Agent Time Distribution Across Iterations (SLM) (\%)}
\label{tab:tab2}
\begin{tabular}{lccc}
\toprule
\textbf{Agent Type} & \textbf{Iter. 1 (\%)} & \textbf{Iter. 2 (\%)} & \textbf{Iter. 3 (\%)} \\
\midrule
Intent UI agent & 13 & 16 & 18 \\
Junior agents (Both) & 7 & 7 & 7 \\
Senior agent & 51 & 52 & 52 \\
Policy agent & 29 & 25 & 23 \\
\bottomrule
\end{tabular}
\end{table}

\begin{table}[t]
\centering
\caption{Agent Time Distribution Across Iterations (gpt-5-nano) (\%)}
\label{tab:tab3}
\begin{tabular}{lccc}
\toprule
\textbf{Agent Type} & \textbf{Iter. 1 (\%)} & \textbf{Iter. 2 (\%)} & \textbf{Iter. 3 (\%)} \\
\midrule
Intent UI agent & 14 & 18 & 21 \\
Junior agents (Both) & 8 & 9 & 8 \\
Senior agent & 53 & 50 & 51 \\
Policy agent & 25 & 23 & 20 \\
\bottomrule
\end{tabular}
\end{table}

\begin{table}[t]
\centering
\caption{Agent Time Distribution Across Iterations (mistral-small3.2:24b) (\%)}
\label{tab:tab4}
\begin{tabular}{lccc}
\toprule
\textbf{Agent Type} & \textbf{Iter. 1 (\%)} & \textbf{Iter. 2 (\%)} & \textbf{Iter. 3 (\%)} \\
\midrule
Intent UI agent & 11 & 12 & 14 \\
Junior agents (Both) & 13 & 13 & 12 \\
Senior agent & 54 & 55 & 53 \\
Policy agent & 22 & 20 & 21 \\
\bottomrule
\end{tabular}
\end{table} 

Table \ref{tab:tab2} \ref{tab:tab3}, \ref{tab:tab4} shows the time consumption for each agent (Senior, Junior and Policy) \textit{w.r.t.} the three metrics (BLEU, METEOR and ROUGE-L). Overall time distribution of the agents is mainly distributed to the Senior agent, which takes up 50\%.This high proportion comes from the tasks of validating and generating the network topology along with their associated weights. Junior agents, tasked with identifying the 5G service type and selecting the appropriate topology, compromise the least portion, consistently consuming the processing time.

The efficiency of the three language models has been evaluated by measuring the total execution time for the complete processing pipeline across three iterations. The experimental results demonstrate that SLM achieves the shortest processing duration, recording 58 seconds for the first iteration, 67 seconds for the second iteration, and 89 seconds for the third iteration. GPT-5-nano has execution times with 80 seconds for the first iteration, 94 seconds for the second iteration, and 107 seconds for the third iteration. Mistral 24B, with 24 billion parameters, requires the longest processing time with 103 seconds for the first iteration, 114 seconds for the second iteration, and 126 seconds for the third iteration.
\section{Conclusion}

This paper has demonstrated a performance evaluation comparing general-purpose LLMs with lightweight, domain-specific SLMs fine-tuned using the LoRA technique. Using metrics such as BLEU, METEOR and ROUGE-L, the results demonstrate that SLMs outperform larger models in both accuracy and time complexity for IBN tasks. 

The future of autonomous networking will be driven by modular Agentic-AI frameworks that incorporate compact, specialized language models suitable for local deployment at the network edge.This shift in paradigm means moving away from relying on general-purpose LLMs and toward a federated, context-aware ecosystem that supports real-time orchestration and closed-loop automation.

\end{document}